\documentstyle[aps,prl,multicol,amssymb]{revtex}
\title{A classical analogue of entanglement}

\author{Daniel Collins$^{1,2}$ and Sandu Popescu$^{1,2}$}

\address{ $^1$H.H. Wills Physics Laboratory,
University of Bristol, Tyndall Avenue, Bristol BS8 1TL, UK
\\ $^2$BRIMS, Hewlett-Packard Laboratories,
 Stoke Gifford, Bristol BS12 6QZ, UK}
\date{6 July 2001}
\begin{document}
\maketitle
\begin{abstract}

We show that quantum entanglement has a very close classical
analogue, namely secret classical correlations.  The fundamental
analogy stems from the behavior of quantum entanglement under
local operations and classical communication and the behavior of
secret correlations under local operations and public
communication.  A large number of derived analogies follow.
In particular teleportation is analogous to the one-time-pad,
the concept of ``pure state'' exists in the classical domain,
entanglement concentration and dilution are essentially classical
secrecy protocols, and single copy entanglement manipulations have
such a close classical analog that the majorization results
are reproduced in the classical setting. This analogy allows one to import
questions from the quantum domain into the classical one, and
vice-versa, helping to get a better understanding of both.  Also,
by identifying classical aspects of quantum entanglement it allows
one to identify those aspects of entanglement which are uniquely
quantum mechanical.

\end{abstract}

\pacs{PACS numbers: 03.67.-a}

\newcommand{\tr}{\mbox{Tr} }
\newcommand{\ket}[1]{\left | #1 \right \rangle}
\newcommand{\bra}[1]{\left \langle #1 \right |}
\newcommand{\amp}[2]{\left \langle #1 \left | #2 \right. \right \rangle}
\newcommand{\proj}[1]{\ket{#1} \! \bra{#1}}
\newcommand{\ave}[1]{\left \langle #1 \right \rangle}
\newcommand{\superop}{{\cal E}}
\newcommand{\unity}{\mbox{\bf I}}
\newcommand{\hilbert}{{\cal H}}
\newcommand{\relent}[2]{S \left ( #1 || #2 \right )}
\newcommand{\banner}[1]{\bigskip \noindent {\bf #1} \medskip}
\newcommand{\I}{{\mathbf I}}
\newcommand{\R}{{\mathbf R}}
\renewcommand{\S}{{\mathbf S}}
\newcommand{\up}{\uparrow}
\newcommand{\down}{\downarrow}

 \begin{multicols}{2}

\section{Introduction}

In his pioneering paper of 1964 \cite{Bell}, J. Bell pointed out
the non-local character of quantum mechanical long-distance
correlations. Since then many novel aspects of non-locality have
been uncovered, such as teleportation \cite{teleport}, super-dense
coding \cite{superdense}, and the capability to reduce the number
of required bits of classical communication for implementing
certain communication tasks (in the so called ``communication
complexity scenario") \cite{complexity}. Furthermore, entanglement
and non-locality are at the core of quantum computation
\cite{qcomp} and its capability of performing computations faster
than any classical computer. An enormous effort has been dedicated
during the last few years to understanding the qualitative and
quantitative properties of non-locality. In effect, quantum
non-locality has become to be considered one of, if not the most
representative aspect of quantum mechanics. Quite surprisingly we
found, as we describe in the present paper, that there exists a
quite close classical analog of quantum entanglement, namely {\it
secret classical correlations}.

Our motivation in looking for a classical analog of quantum
entanglement is two-fold. Firstly, such an analogy allows us to
identify aspects of quantum entanglement which were hitherto
considered to be purely quantum but which are in fact not quantum
at all. Indeed, all those aspects of entanglement which are common
with the classical analog, are not of a quantum nature. As a
corollary we also get a better understanding of what are the true
quantum features of quantum entanglement. Secondly, this analogy
allows one to transfer questions from quantum entanglement to the
classical domain (classical information cryptography) and
vice-versa and thus lead to a better understanding of both
subjects. In fact, the inspiration for our paper stems from the
work of N. Gisin and S. Wolf \cite{Wolf} which asked if there is a
classical analog of bound entanglement.

The analogy we suggest is summarized in the following table:

\begin{center}
\begin{tabular}{l l l}
quantum & \hspace{0.5in} & secret classical \\ entanglement & &
correlations \\ & & \\

quantum & & secret classical \\
communication & & communication \\

& & \\

classical & & public classical \\
communication & & communication \\

& & \\

local actions & & local actions

\end{tabular}

 \end{center}

Thus we suggest that a classical analog of a pair of entangled
particles is that of one sample of two secret, correlated, random
variables (one at each remote party). Here by secret communication
we mean communication through a channel to which an eavesdropper
has no access. By public communication we understand communication
through a channel to which an eavesdropper has full access (can
hear everything), but cannot alter the messages sent, nor
introduce new messages. Finally, in the quantum context by local
actions we understand subjecting the q-bits to unitary evolutions
as well as to measurements and other non-unitary evolutions. The
classical analog of unitary transformations is that of replacing
the value of the original random variable by some new value
related to the old one by a one-to-one function, while the analog
of the case of quantum non-unitary evolutions is that of
transformation by non bijective functions. \footnote{Note that
when we replace the original value of the random variable by
another via a non-bijective function, we consider that we actually
erase the original information, so information is lost. This is
completely analogous to what happens in the quantum case. Of
course, one may argue that in neither case information is lost.
For example, in the non-collapse interpretations of the quantum
case all we have is an entanglement of the measured system with
the measuring device; this entanglement however involves so many
degrees of freedom that it cannot be reversed. Similarly, erasing
say pencil markings from a paper still preserves the original
information in some subtle arrangement of the graphite granules
mixed with bits of paper and erasing gum, but this involves so
many degrees of freedom that the original information cannot be
recovered.}

The main idea of this analogy is that as with quantum
entanglement, secret classical correlations act as a (fungible)
{\it resource} and obey a ``second law of thermodynamics"
principle - the amount of secrecy doesn't increase under LOPC
(local actions and public communication).

The modern paradigm is that of quantum non-locality as a {\it
resource} as we describe below .
\begin{itemize}
\item
Non-local correlations between two or more remote parties can be
created by quantum communication, i.e. by sending quantum
particles (q-bits) from a common source to the parties, or from
one party to another.

\item {\it Second law of thermodynamics}:
The amount of non-locality between the remote parties cannot be
increased by local actions and/or classical communication (LOCC).

Indeed, one can view this statement as the very {\it definition}
of what non-locality is.

The above version of the second law can be further extended to
allow for quantum communication, catalysis, etc.. For example
\cite{Lo2} `` By local actions, classical communication and
exchange of n q-bits, the amount of non-locality between remote
parties cannot be increased by more that n e-bits".

\item The remote parties can, by local actions
and classical communication, transform non-locality from one form
into another.

For example, suppose two parties, Alice and Bob, have a large
number of pairs of particles, each pair in some pure,
non-maximally entangled state. By appropriate actions they can end
up with a smaller number of pairs in maximally entangled states
\cite{Concentration}, \cite{Lo2}. In effect, at least in the case
of bi-partite pure states, non-locality is absolutely fungible -
any form can be transformed into any other, and the transformation
is reversible. Thus it doesn't really matter in which form the
parties are supplied with non-locality, they can always convert it
into the form which is required for implementing the specific task
(for example teleportation) they want to do.

\item
Non-locality is consumed for producing useful tasks
(teleportation, super-dense coding, remote implementation of joint
unitary transformations \cite{Unitary}, etc.).
\end{itemize}

As with quantum non-local correlations, secret correlations are
also a resource.
\begin{itemize}
\item
 Secret correlations can be
established between remote parties by secret communication.

\item ``Second law of thermodynamics": The amount of secret
correlations cannot be increased by local actions and/or public
communication (LOPC)\footnote{In everyday practice, secret
messages are exchanged by public communication by so called
``public key distribution" protocols. We do not consider here this
case since these are only pseudo secret messages - their secrecy
is based on encoding which is difficult to decode due to
computational complexity; in principle however an eavesdropper
could decode the message.}.

In fact, as with the case of non-locality, we can take this law to
be the very definition of the amount of secret correlations, i.e.
the amount of secret correlations between remote parties is that
part of their correlations which cannot be increased by local
actions and public classical communication.

The above version of the second law can be further extended to
allow for secret communication, catalysis, etc. For example
 `` By local actions, public communication and
exchange of n secret bits, the amount of secret correlations
between remote parties cannot be increased by more that n secret
correlation bits".

\item The remote parties can, by local actions
and public communication, transform secret correlations from one
form into another.
\item
Analogous to entanglement, secret correlations are a fungible
resource - they can be stored, transformed from one form into
another, and can be consumed to perform useful tasks, such as
secret communication via the one time pad \cite{onetimepad}.
\end{itemize}

The possibility of transforming secret correlations from one form
into another enables us, similarly to the case of quantum
correlations, to obtain a {\it quantitative} description of
secrecy.

In the bi-partite case, the analogy is now obvious:

Shared, undirected resources:
\begin{center}
\begin{tabular}{l l l l}
$e-bit_{AB}$ & \hspace{0.7in} & shared secret $bit_{AB}$ & \hspace{0.2in}
\end{tabular}
\end{center}

Directed resources:

\begin{center}
\begin{tabular}{l l l}
$qubit_{A \rightarrow B}$ & \hspace{0.3in} & secret
$bit_{A \rightarrow B}$ \\
& & \\
classical $bit_{A \rightarrow B}$ & & public classical
$bit_{A \rightarrow B}$
\end{tabular}
\end{center}

The situation of multi-partite secret correlations is more
complicated, as is the situation of multi-partite entanglement. It
is now clear that there are many different, irreducible, types of
multi-partite entanglement \cite{multi1}, \cite{multi2};
this is also the case for secret correlations.

At this point it is legitimate to ask what is the role of secrecy.
That is, why do we consider {\it secret} classical correlations to
be the analogue of entanglement and not simply {\it any} classical
correlations. There are two main reasons. First of all, while such
an analogy is certainly possible, it would be rather
uninteresting. Indeed, one of the main aspects of manipulating
entanglement is that there is a way in which the different parties
may communicate (classical communication) which doesn't increase
the amount of entanglement. Similarly in the case of secret
classical correlations, public communication doesn't increase the
amount of secrecy. In the case of arbitrary classical correlations
however there is no way in which the remote parties could
communicate and not increase the correlations. So when trying to
build an LOCC (``local operations and classical communications")
analog in the case of arbitrary classical correlations we have no
choice but to completely eliminate the communication, which leads
to a very uninteresting situation.

The second reason is far more profound. Consider for example two
parties, Alice and Bob who share, say, a maximally entangled state
$\ket{\Psi}={1\over{\sqrt2}}(\ket{0} \ket{0} + \ket{1} \ket{1})$.
Suppose now that Alice
and Bob ``degrade" the state by ``erasing" the entanglement. They
can do this {\it in a minimal way} by, say, Alice randomizing the
phase of her basis state vectors $\{ \ket{0}, \ket{1} \}$.
Then Alice and Bob will be left with a mixture of
${1\over{\sqrt2}}(\ket{0} \ket{0} + \ket{1} \ket{1} )$ and
${1\over{\sqrt2}}(\ket{0} \ket{0} - \ket{1} \ket{1} )$ with equal
probabilities. This
mixture contains no entanglement (it is equivalent to an equal
mixture of $\ket{0} \ket{0}$ and $\ket{1} \ket{1} $) but contains
secret correlations
between Alice and Bob. Thus secret correlations are in fact very
closely related to entanglement.

The analogies described above are the ``fundamental" analogies.
From them follow an entire set of derived analogies. We would like
to emphasize however that it is only the fundamental analogies
(such as the behavior under LOCC/LOPC) which have truly deep
significance and that one shouldn't expect the derived
analogies to be very close (though many of them are). Derived
analogies are summarized in the following table:

\begin{center}

\begin{tabular}{l l l}

teleportation & \hspace{0.5in} & one-time pad  \\

& & \\

entanglement & & secret correlation \\
concentration & & concentration \\

& & \\

entanglement & & secret correlation \\
dilution & & dilution \\

& & \\

entanglement & & classical privacy \\
purification & & amplification \\

& & \\

single copy & & single copy \\
transformations & & transformations \\

& & \\

probabilistic single & & probabilistic single \\
copy transformations & & copy transformations \\

& & \\

catalytic & & catalytic \\
transformations & & transformations \\

 & & \\

bound & & bound \\ entanglement & & information ? \\

\end{tabular}

\end{center}

\section{Quantum states and classical analogues}

\label{sec-states}

In the previous section we suggested that classical secret
correlations are a good analog for quantum entanglement. Again,
the basis of the analogy is the similar behavior of secret
correlation and quantum entanglement under LOPC/LOCC. To make the
analogy more detailed and to obtain the ``derived" analogies
mentioned above we need to define more precisely the analogy
between quantum states and secret correlations.

Consider two remote parties, Alice and Bob. A general quantum
state is described by a density matrix $\rho_{AB}$ or,
equivalently, by a pure state $\Psi_{ABE}$ in which A and B are
entangled with a third party, the ``environment". The classical
equivalent of the general quantum state is a probability
distribution $P(X_A, X_B,X_E)$ where $X_A$, $X_B$ and $X_E$ are
random variables known to Alice, Bob and Eve (the eavesdropper)
respectively. One copy of a quantum state $\Psi_{ABE}$ corresponds
to one sample of the probability distribution $P(X_A, X_B,X_E)$.

A quantum bi-partite pure state can always be written in the
Schmidt basis \cite{Schmidt} as

\begin{equation}
\ket{\psi}_{AB} = \sum_i \sqrt{p_i} \ket{i}_A \ket{i}_B.
\end{equation}

If Alice and Bob measure their particles in the Schmidt basis
 then they get correlated random variables,
$X_A$ and $X_B$, which come according to the distribution $p(X_A =
i, X_B = j) =  \delta_{ij} p_i$. In other words, they both get the
same sample from a random variable $X \sim \{ p_i \}$.
Furthermore, the values of $X_A$ and $X_B$ are secret - there is
no third party E who knows them.   We propose classical
distributions of this form as the classical ``pure'' state. That
is, a bi-partite classical pure state is a distribution
\begin{equation}
p(X_A = i, X_B = j, X_E=k) =  \delta_{ij} p_i \tilde
P(E_k)\label{purestatedefinition}
\end{equation}
 where
$\tilde P(E_k)$ is the distribution of eavesdropper's variable
$X_E$ and it completely irrelevant, except for the fact that it is
completely uncorrelated to the distribution of $X_A$ and
$X_B$\footnote{Note that quantum mechanically in order to say that
the state of Alice and Bob is pure we don't need to specify that
the state of Alice, Bob and the Environment is of the form
$\ket{\psi}_{ABE}=\ket{\psi}_{AB}\ket{\tilde\psi}_{E}$, but it is
enough to know the state $\rho_{AB}$ of Alice and Bob alone. On
the other hand, the classical correlations of Alice and Bob alone
do not allow us to know if Eve is, or is not, correlated with
Alice and Bob, therefore we must always describe the full state of
Alice, Bob and Eve.}.  Strictly speaking, we propose
(\ref{purestatedefinition}) as the classical analogue of the
pure state Schmidt decomposition, and any classical state
which is locally equivalent, ie. can be transformed into the
above form by local, one-to-one
mappings (the equivalent of local unitaries) we consider
to be a pure state.

Another interesting case is that of distributions of the form
$p(X_A = i, X_B = j, X_E=k) = P(X_A=i, X_B=j)\tilde P(E_k)$ in
which E is completely uncorrelated with A and B, but A and B are
not completely correlated with each other.  Such a distribution is
obtained when Alice and/or Bob measure a quantum pure state in
some other basis than the Schmidt one. Such a distribution has
some characteristics of a pure state and some characteristics of a
mixed state. We will discuss in more detail this case in section
\ref{sec-pure}.

For more than two parties the analogue of a density matrix
$\rho_{ABC...}$ is a probability distribution
$P(X_A,X_B,X_C,...)$.  It is not yet clear to us what the general
analogue of a multi-partite pure state is. This is due, in part,
to the fact that for multi-partite states the analog of the
Schmidt decomposition is far more complicated.  We shall give some
multipartite results in section~\ref{sec-Multi}.

\section{Teleportation and the One Time Pad}

The first ``derived" analogy is probably the most striking of all.
The fundamental quantum communication protocol that is
teleportation turns out to be analogous to the fundamental secret
communication protocol, the one-time pad
\cite{teleportationanalog}.

Alice begins with the qubit (secret bit) to be sent, which may be
entangled (secretly-correlated) with any number of other particles
(bits). She does a Bell measurement (addition modulo 2) on the
qubit (secret bit) to be sent and the qubit (bit) of resource she
holds. She then sends the outcome (result) of this operation as a
classical bit (public bit) to Bob.  He then does a conditional
unitary (bit flip) upon his part of the e-bit (shared secret bit).
Bob now holds the qubit (secret bit) Alice was sending him.

The necessary and sufficient resources are given by:
\begin{equation}
1 e-bit_{A B} + 2 classical~ bits_{A \rightarrow B} \Rightarrow 1
qubit_{A \rightarrow B}
\end{equation}
\begin{equation}
1 shared~ secret~ bit_{A B} + 1 public~ bit_{A \rightarrow B}
\Rightarrow 1 secret~ bit_{A \rightarrow B}
\end{equation}
By necessary we mean that, if we were to try to do the
teleportation with less than 1 e-bit - by using a less than
maximally entangled state for example - the teleportation will not
give a perfect output, and the classical information will give
some information about the qubit we are sending.  If we try to use
a less than completely correlated shared secret bit to send a
secret bit then Eve gets some information about the secret bit.
The resources are sufficient since we can achieve the operations
using them.

Note that the resources are used up in the process: once we have
used an e-bit (shared secret bit) to send a qubit (shared secret
bit) we cannot reuse it. Quantum mechanically this is obvious,
since the original maximally entangled state is destroyed by
Alice's measurement. Classically however Alice and Bob do not lose
their correlated bits -  Alice and Bob need not erase or
physically modify in any way their original correlated bits but
just use them for some mathematical operations. What is lost
however is the secrecy of these bits - they cannot be reused.

Furthermore, it is obvious to see that the one-time pad secret
communication can be used to implement the analog of teleportation
of entangled states and of entanglement swapping.

Finally, let us note an important fact. Quantitatively the amount
of resources in the classical and quantum cases are similar but
not identical: but we need 2 classical $bits_{A \rightarrow B}$ to
send 1 qubit, whereas only 1 public $bit_{A \rightarrow B}$ to
send 1 secret bit.

\section{Entanglement and Secret Correlation Manipulations - Single Copy}

\label{purestatesection}

The ability to manipulate entanglement, i.e. transforming
entanglement from one form into another by local actions and
classical communications is one of the most important aspects of
entanglement. This leads to elevating entanglement to the status
of a (fungible) resource: to a large extent it doesn't matter in
which form entanglement is supplied, we can transform it into the
specific form we need for different applications, very much as
say, transforming the chemical energy stored in coal into
electrical energy for use in electric engines. Similarly one can
imagine that Alice and Bob are supplied with secret correlations
in some given form, i.e. according to some specific probability
distribution, and they want to obtain secret correlations obeying
a different probability distribution. We find that the quantum and
classical scenarios are in very close analogy.

In this section we treat the case of bi-partite pure state single
copy manipulations. In the quantum context this means that the two
parties, Alice and Bob, share a single pair of particles in some
pure state $\ket{\Psi}_{AB}$. In the classical context, Alice and Bob
share a single sample of a classical pure-state
(\ref{purestatedefinition}).

In the case of a single copy, entanglement is not a completely
interconvertible resource (as it is in the case of many copies
(see section \ref{sec-concentration})), but many more restrictions
apply.

For bipartite pure quantum states, it is possible to turn one
state into another {\it with certainty} if and only if a certain
set of conditions, collectively known as majorization, holds
\cite{Lo}, \cite{Nielsen}.  We here show that for
classical secret pure states, the transformation is possible if
and only if an analogous condition holds.

Quantum mechanically the majorization condition is the following.
Consider two quantum pure states $\ket{\psi}_{AB}$ and
$\ket{\phi}_{AB}$, written in their Schmidt bases
\begin{equation}
\label{eqn-psi}
\ket{\psi}_{AB} = \sum_i \sqrt{p_i} \ket{i}_A \ket{i}_B,
\end{equation}
\begin{equation}
\label{eqn-phi}
\ket{\phi}_{AB} = \sum_i \sqrt{q_i} \ket{i}_A \ket{i}_B,
\end{equation}
with the squared Schmidt coefficients $p_i$ and $q_i$ arranged in
decreasing order, $p_1 \ge p_2 \ge ...$ and $q_1 \ge q_2 \ge ...$.
The vector $\vec{q} = \{ q_i \}$ is said to majorize the vector
$\vec{p} = \{ p_i \}$ iff
\begin{equation}
\sum_{i=1}^k q_i \ge \sum_{i=1}^k p_i~~~ \forall k.
\end{equation}
$\ket{\phi}_{AB}$ is said to majorize $\ket{\psi}_{AB}$ iff
$\vec{q}$ majorizes $\vec{p}$. The transformation $\ket{\psi}_{AB}
\mapsto \ket{\phi}_{AB}$ is possible with certainty if and only if
$\ket{\phi}_{AB}$ majorizes $\ket{\psi}_{AB}$ \cite{Nielsen}.
(Note that it is the final state which must majorize the starting
one.)

For classical secret correlations, suppose Alice and Bob begin with
an arbitrary classical bipartite pure state, which we may write as
\begin{equation}
\label{eqn-X}
p(X_A = i, X_B = j, X_E=k) =  \delta_{ij} p_i \tilde P(E_k).
\end{equation}
Their task it to produce some other state,
\begin{equation}
\label{eqn-Y}
p(Y_A = i, Y_B = j, Y_E=k) =  \delta_{ij} q_i \tilde P'(E_k).
\end{equation}
We shall prove that they can do this iff
$\vec{q}$ majorizes $\vec{p}$.  However, to understand what is going
on, let us first consider a simple example which has all the important
features.  The quantum version was first considered in \cite{Lo}.

Suppose Alice and Bob share one sample of the classical pure state
$X$, where
\begin{equation}
p_1 = p_2 = p_3 = \frac{1}{3},
\end{equation}
and they would like to turn it into a sample of the pure state
$Y$, where
\begin{equation}
\label{sharedsecretbit}
q_1 = q_2 = \frac{1}{2}.
\end{equation}

A probabilistic method (analogous to the procustean method for the
quantum case\cite{Concentration}) is for Alice to send message
$m_1$ (which means ``OK") if X is 1 or 2, and to send message $m_2$
(which means ``not OK") if X is 3.
If message $m_1$ is sent then Alice and Bob
keep their sample, and they now have a shared secret random
variable of the form $Y$. Indeed, in this case Eve only knows that
the  value of the secret variable is either 1 or 2 but she doesn't
know which one - Alice and Bob's data is therefore still perfectly
secret, and it is now either 1 or 2 with probability $1/2$. If
message $m_2$ is sent then the procedure failed and Alice and Bob
have to throw away their sample. The reason is that Eve, who
monitors the public communication, learns that Alice and Bob's
variable is equal to 3, and there is no more Alice and Bob can do.

The above method works with probability $\frac{2}{3}$. Can Alice
and Bob do better? The second distribution majorizes the first,
since $\frac{1}{2} \ge \frac{1}{3}$, $\frac{1}{2} + \frac{1}{2}
\ge \frac{1}{3} + \frac{1}{3}$ and $\frac{1}{2} + \frac{1}{2} + 0
\ge \frac{1}{3} + \frac{1}{3} + \frac{1}{3}$. Thus, according to
the majorization theorem we shall shortly prove, there exists a
method which works with certainty. The protocol for achieving this
goes as follows. Alice reads the value of X.  If it is 1, she
flips an unbiased coin which tells her to send message $m_1$ or
$m_2$ with equal probability. If $X = 2$ she flips an unbiased
coin to send $m_2$ or $m_3$, and if $X = 3$ she flips an unbiased
coin to send $m_1$ or $m_3$.  She then publicly sends the message,
so that everyone can read it.  If $m_1$ is sent, Eve knows that X
is 1 or 3 with equal probability. If $m_2$ is sent, Eve knows that
X is either 1 or 2, with equal probability. And if $m_3$ is sent,
Eve knows that X is 2 or 3 with equal probability. Now Alice and
Bob just have to do a simple relabelling of X to produce Y.  If
$m_1$ is sent, they both do $1 \mapsto 1, 3 \mapsto 2$.  If $m_2$
was sent they do $1 \mapsto 1, 2 \mapsto 2$.  If $m_3$ is sent
they do $2 \mapsto 1, 3 \mapsto 2$. Whatever message was sent, Y
is now a shared random variable which is (as far as Eve is
concerned) a shared secret bit of the form (\ref{sharedsecretbit}).

Now we shall look at the general case.  For which pure states X
and Y is it possible to turn a single sample of X into a single
sample of Y?  Consider the most general possible protocol.  We
assume that Alice, Bob and Eve all know the protocol\footnote{if
Alice and Bob had a secret protocol, this would be like having an
additional shared random variable, whose different outcomes told
them which protocol to use.  Thus they would have an additional
resource. Here we insist they have only one shared resource, X.}.
Alice and Bob start by having a single sample of the pure state X.
They each have also access to some local source of secret
randomness - they may each throw dice. Of course, Alice knows only
the outcomes of her dice and Bob of his. During the protocol Alice
and Bob may publicly communicate, perhaps in many rounds, with
each message determined by X, the public messages already sent,
and by the results of the local dice.  At the end of the protocol
there will be some total public message which consists of all the
messages that were exchanged by Alice and Bob.  All three parties,
Alice, Bob and Eve know this total message.  In addition, Alice
and Bob know the value of X (which is common to both of them since
the state is pure), and each of them knows the outcomes of his/her
own dice. Based on all this knowledge Alice and Bob must decide on
the values of $Y_A$ and $Y_B$. Formally, we can write
\begin{equation}
Y_A=f_A(X_A, m, d_A)
\end{equation}

\begin{equation}
Y_B=f_B(X_B, m, d_B)
\end{equation}
where by $m$ we denote the total message, and by $d_A$ and $d_B$
we denote the outcome of all Alice's and Bob's dice.

The above procedure can be simplified. Since we begin with a pure
state, $X_A=X_B=X$. Furthermore, since we want to end with a pure
state, we require $Y_A=Y_B$. This requirement implies that $Y_A$
and $Y_B$ cannot depend on the outcome of the dice $d_A$ and
$d_B$. Also given the initial value X and the message m, Alice and
Bob must perform the same function $f$. Thus we get
\begin{equation}
Y_A=Y_B=f(X, m).
\end{equation}

Furthermore, since Bob's actions may not depend on the outcomes of
his dice but only on X and m, for every procedure which involves
many rounds of communication between Alice and Bob, we can
formulate an equivalent procedure in which the total message is
entirely generated by Alice - she could simply throw all dice
herself - and then communicate the message in a single
transmission to Bob.

Let us now formalise this procedure for turning X into Y.

Alice looks at $X = x_i$, which occurs with probability $p_i$. She
then throws a biased dice which tells her to send message $m_j$
with some probability $p(m_j|x_i)$ which depends upon $x_i$. She
then publicly announces $m_j$. Alice and Bob now follow the
instructions in the message, which say to do $x_i \mapsto y_k(x_i,
m_j)$.  Forgetting what X is (ie. summing over $x_i$) this gives
them some joint distribution for $y_k$ and $m_j$, $p(y_k, m_j)$.
Since Alice and Bob want $y_k$ to be secret from Eve, who knows
only the protocol and the message, this distribution must
factorise: $p(y_k, m_j) = p(y_k) p(m_j)$. $p(y_k)$ is the final
distribution, and so we want $p(y_k) = q_k$ (the distribution of
Y).

This secrecy procedure can be thought of as a single party problem,
which goes as follows.  We begin with a sample from X, which
occurs with probability $p_i$.  We may look at the sample, and
then roll some dice which gives outcome $m_j$ with probability
$p(m_j | x_i)$.  We then perform the map $x_i \mapsto y_k(x_i,m_j)$.
We then forget what X is, which gives some joint distribution
for $y_k$ and $m_j$, $p(y_k,m_j)$.  We desire this distribution
to factorise, $p(y_k,m_j) = p(y_k) p(m_j)$, and that $p(y_k) = q_k$.
Note that this single party procedure is not a secrecy procedure,
however it is possible iff the above secrecy transformation is.

To find for which $p_i$ and $q_k$ this single party problem is
possible, and thus to find for which $p_i$ and $q_j$ the secrecy
transformation is possible, we shall look at the time reversed
problem.  This goes as follows. We start with a sample from Y,
which occurs with probability $q_k$. We then roll dice, which give
outcome $m_j$ with probability $p(m_j)$, independent of the
outcome of Y.  This gives a joint distribution $p(y_k, m_j) = q_k
p(m_j)$.  Now we must do the inverse of the map $x_i \mapsto
y_k(x_i, m_j)$ to turn our Y into an X. If the map is one-to-one,
and hence invertible, this will give us a distribution $p(x_i,
m_j)$. Like any joint distribution, this can be written as $p(x_i,
m_j) = p(x_i) p(m_j | x_i)$.  If we now forget the value of Y and
of $m_k$, we get a new distribution for X, $p(x_i)$.  We desire
$p(x_i) = p_i$.  If the map is many-to-one, then we can give it a
probabilistic inverse which is a ``one-to-many'' map where the
probabilities of getting various $x_i$'s given any particular
$y_k$ are given by the relative frequencies of the $x_i$'s when
$y_k$ is produced in the forward time protocol. This probabilistic
one-to-many map can be replaced by a probabilistic choice of
several one-to-one maps, which will have the same effect upon the
protocol since we forget which map we did at the end. Thus in the
reversed time single party problem, we need only consider maps
which are one-to-one.  This also applies to the forward time
single party problem, and to the forward time secrecy protocol: we
only need consider maps which are one-to-one, ie. permutations.

As explained above, if we find the
conditions for which the reversed time single party problem is
possible, we will have the conditions for which the forward time
secrecy transformation is possible.  Physically, this time reversed
single party problem goes as follows.  We begin with a
ball in some box according to the distribution $q_i$. We do not
know which box the ball is in, and are not allowed to look to see
where it is.  We then apply some shuffle
(one-to-one relabelling) to the boxes, choosing which shuffle to make
according to a distribution, $p(m_j)$, which we may choose.  We then
forget which shuffle we did, and look at the new distribution of
the balls, $p_i$.  The question is for which $q_i$ and $p_i$ is
this possible? Clearly $p_i$ should be more random than $q_i$.
This is a well-known  problem, and is the context in which
majorization appears in classical physics.  The answer is that it
is possible iff $\vec{p}$ majorises $\vec{q}$.  Intuitively this
is easy to see, and the proof can be found, for example,
in\cite{Wehrl}.

Above we have proved the majorization result in the classical
context by using arguments referring solely to the classical
context. We could have used however the known results for quantum
entanglement manipulation to prove the classical ones. The reason
is as follows. On one hand, it was found out that transforming
pure quantum states (with certainty) from one into another
involves only actions and measurements in the Schmidt
decomposition basis.  These actions do not involve phases, but are
simply classical actions upon the basis, which are performed
coherently to make a quantum evolution.  One could, however,
imagine starting by measuring the quantum state in the Schmidt
basis, and then performing the corresponding classical actions and
measurements upon the state.  This transforms one classical state
into another, and will not give Eve any knowledge about the state
since the quantum procedure did not entangle the quantum state
with the environment. Thus, if we can transform with certainty a
quantum pure state $|\Psi\rangle$ (\ref{eqn-psi}) into a quantum
pure state $|\Phi\rangle$ (\ref{eqn-phi}), we can also transform
with certainty $X$ (\ref{eqn-X})
the classical pure state equivalent of $|\Psi\rangle$, into $Y$
(\ref{eqn-Y}), the classical pure state equivalent of $|\Phi\rangle$.

To prove the reverse, that is, that $X$ can be transform with
certainty into $Y$ only if the quantum analogs can be transformed
from one into the other, we note that we can turn any classical
transformation of pure states into a quantum one, simply by
applying the classical operations coherently, and performing the
quantum actions in the Schmidt basis.  Thus there cannot be any
classical procedure which does better than the optimal quantum
one.  So the classical transformation is possible iff the quantum
one is. \footnote {Note however that although we can use the
quantum result to prove the classical one, we cannot use the
classical result to prove the quantum result. The reason is that
although we can turn any classical transformation into a quantum
one, we cannot generate this way all possible quantum protocols -
indeed, they may involve phases outside the Schmidt basis.}

\section{Probabilistic Single Copy Manipulations}

It may not be possible to transform a single copy of a resource
from one form into another with certainty, but it may be possible
to do it with some probability.  What is the largest probability
with which this can be done?  For quantum states, the problem was 
considered in \cite{Lo}, \cite{Vidal}, and the general answer is
given \cite{Vidal} in the simple form:
\begin{equation}
\min_k \frac{1 - \sum_{i=1}^k p_i}{1 - \sum_{i=1}^k q_i}.
\end{equation}
We shall now show that for classical secret states, the answer is
the same.

As for the non-probabilistic transformations, we may simplify the
most general protocol, which then goes as follows. Alice first
looks at her sample which comes according to the distribution
$p(x_i)$.  She then chooses a message $m_j$ according to
$p(m_j|x_i)$.   Most of the possible messages will be ones for
which the transformation succeeds: these must say to do a
one-to-one map\footnote{ There is no loss in generality in
forgetting about the many-to-one maps, for the same reasons as in
the non-probabilistic manipulations.}  $X \mapsto Y$.  The other
messages say ``fail": for these it does not matter what
transformation we do, and it does not help to send more than one
``fail" message.  So we may assume we have only one ``fail"
message, $m_{fail}$, which says to do $x_i \mapsto y_1$. Alice and
Bob then do $x_i \mapsto y_k(x_i, m_j)$ according to the message.
This gives them a distribution $p(y_k, m_j)$.  In the case they
succeed, this distribution must factorise:
\begin{equation}
p(y_k, m_j) = \left\{ \begin{array}{ll} p(y_k) p(m_j) &  \mbox{for
j $\neq$ ``fail"} \\ \delta(y_k=1) p(m_{fail}) & \mbox{for j =
``fail"}
\end{array} \right.
\end{equation}
By defining $p(success) = \lambda$, so that $p(m_j)=\lambda p(m_j|
success)$ for $j \neq ``fail"$ and $p(m_{fail})=1-\lambda$ and by
requiring $p(y_k) = q_k$ (so that the protocol succeeds) we
obtain:
\begin{equation}
p(y_k, m_j) = \left\{ \begin{array}{ll} \lambda q_k p(m_j|
success) &  \mbox{for j $\neq$ ``fail"} \\ (1- \lambda)
\delta(y_k=1) & \mbox{for j = ``fail"}   \end{array} \right.
\end{equation}

The time reversed, single party version of this problem
is to start by flipping a coin (H/T) with
probabilities $(\lambda, 1-\lambda)$.  We look at the result, and
if it is T we start with $y_k = 1$, send a message $m_{fail}$,
and are allowed to do anything (including probabilistic things)
to transform $Y \mapsto X$.  If the coin is H we get a sample $y_k$
according to $p(y_k)=q_k$, but do not know which sample we get.  We
then pick some message according to $p(m_j)$, and do the corresponding
shuffle $y_k \mapsto x_i$.  This gives some distribution
$p(x_i, m_j)$.  Finally, we forget whether the coin
was H or T, and also which message was sent.  This then gives us
$p(x_i)$, which we would like to be $p_i$.  Our aim is, for a given
$q_k$ and $p_i$, to find the maximal $\lambda$ for which this is
possible.  This problem is closely related to the one where majorization
first appeared in classical physics, and the maximal value of
$\lambda$ is as given at the start of this section.  Once again the
quantum and classical pure state manipulations are possible under
the same conditions.

\section{Catalysis of Single Copy Transformations}

\label{sec-catalysis}

There is an interesting entanglement transformation called catalysis
\cite{Jonathan} which transfers easily to the classical case.
Suppose we begin with some pure state
\begin{equation}
\ket{\psi}_{AB} = \sum_i \sqrt{p_i} \ket{ii}_{AB},
\end{equation}
and wish to produce, using LOCC, the state
\begin{equation}
\ket{\phi}_{AB} = \sum_j \sqrt{q_j} \ket{jj}_{AB}.
\end{equation}
This is possible\cite{Nielsen} iff $q_j$ majorizes $p_i$.
There are, however, states such that $q_j$ does not majorise $p_i$,
but where catalysis is possible.  That is, where Alice and Bob
cannot perform
\begin{equation}
\ket{\psi} \mapsto \ket{\phi},
\end{equation}
but if Alice and Bob share an additional pure state,
\begin{equation}
\ket{\chi}_{AB} = \sum_k \sqrt{r_k} \ket{kk}_{AB},
\end{equation}
then they are able to perform, with certainty, the
transformation
\begin{equation}
\ket{\psi} \ket{\chi} \mapsto \ket{\phi} \ket{\chi}.
\end{equation}
This is, quite simply, because for the tensor product system,
the majorization condition holds.  $\ket{\chi}$ acts as a
catalyst.  It enables the transformation of $\ket{\psi}$ into
$\ket{\phi}$, but is not consumed in the process.  One
example of such a catalysis is transforming the quantum state
whose squared Schmidt coefficients are
\begin{equation}
p_1 = 0.4; p_2 = 0.4; p_3 = 0.1; p_4 = 0.1
\end{equation}
into the quantum state
\begin{equation}
q_1 = 0.5; q_2 = 0.25, q_3 = 0.25,
\end{equation}
using the catalyst
\begin{equation}
r_1 = 0.6; r_2 = 0.4.
\end{equation}

The classical analogue of this process follows immediately.
That is, Alice and Bob may wish to turn the classical pure
state defined by $p_i$ into the classical pure state defined
by $q_j$, using LOPC.  This is only possible, as we showed in section
\ref{purestatesection}, when $q_j$ majorises $p_i$.
However there are cases when this is not possible, but if
they also have a sample of the classical pure state $r_k$,
then they can achieve the transformation
\begin{equation}
P \otimes R \mapsto Q \otimes R
\end{equation}
with certainty.  The sample R is not revealed or altered by this
process, and can be subsequently used independently elsewhere. As
far as we know, this classical secret correlation catalysis has
not been previously considered.

\section{Shuffling with Catalysis}

Another classical catalysis problem which has not (to our
knowledge) been considered before is the single party, time
reversed version\footnote{see section \ref{purestatesection} for
the meaning of the single party, time reversed version of the
classical pure state transformation.} of the classical pure state
catalysis discussed in the previous section.  We call this
``shuffling catalysis''.
We emphasize that this shuffling catalysis has,
in itself, nothing to do with secrecy or secret correlations.
However, it is possible to perform this shuffling catalysis iff
the classical pure state catalysis is possible.  Recalling
(from section \ref{purestatesection} that the majorization
conditions are easier to prove in the shuffling scenario than
in the classical secret correlation scenario, studying shuffling
catalysis may help in finding exactly when classical secret
correlation (and, by analogy, entanglement) catalysis is possible.

We state the problem of shuffling catalysis to make the idea
clear. Suppose we have a sample from a distribution $q_j$ and wish
to turn it into a sample from a distribution $p_i$.  We are not
allowed to look at the sample to see what it is, we can only throw
dice whose probabilities (which we choose) are independent of
which sample we have.  We then make some permutation (shuffle)
upon the outcomes, which suffle decided by the dice, and finally
forget which one we did.  As mentioned in section
\ref{purestatesection}, this ``shuffling'' is possible iff $q_j$
majorizes $p_i$. There are, however, distributions where $q_j$
does not majorize $p_i$, and so cannot be turned into it directly,
but where we can perform catalysis.  This means that we can take a
sample from a third distribution $r_k$, such that $q_j \otimes
r_k$ majorizes $p_i \otimes r_k$, and then roll an independent
dice and permute the possible outcomes of the tensor product
distribution to turn $q_j \otimes r_k$ into $p_i \otimes r_k$.
This catalysis is possible iff we can use $r_k$ to turn the shared
secret correlation pure state $p_i$ into to the pure state $q_j$.
Thus an example of this shuffling catalysis is the example given
in section\ref{sec-catalysis}.

\section{Pure State Concentration and Dilution}

\label{sec-concentration}

For many copies of a bipartite pure state, entanglement is a
completely fungible resource.  It can be converted from one form
to another reversibly.  Thus we can quantify the amount of
entanglement by a single number, the entropy of entanglement. We
shall show that the same is true for classical pure bipartite
states.  That is, for such states, secret correlations are a
completely fungible resource.  They can be converted from one form
to another reversibly, and can be quantified by a single number,
the entropy of secrecy.

We define the entropy of entanglement for a quantum pure state,
$E(\ket{\psi}_{AB})$ as
\begin{equation}
E(\ket{\psi}_{AB}) = -\sum_i p_i\log p_i
\end{equation}
where $p_i$ are the squares of the Schmidt coefficients.

The physical meaning of the entropy of entanglement is the
following. When Alice and Bob share a large number $N$ of copies of
some arbitrary pure state $\ket{\psi}_{AB}$, they can convert
them, in a {\it reversible way}, using only local operations and
classical communication into a number $K$ of copies of the maximally
entangled state
\begin{equation}
\ket{\psi_s}_{AB} = \frac{1}{\sqrt{2}} ( \ket{11}_{AB} +
\ket{22}_{AB} )
\end{equation}
where
\begin{equation}
 \frac{K}{N} \rightarrow E(\ket{\psi}_{AB})
\end{equation}
as $N \rightarrow \infty$. That is, the entropy of entanglement
represents the yield of singlets per copy of the original state
$\ket{\psi}_{AB}$.  The operation of converting the states
$\ket{\psi}_{AB}$ into maximally entangled states is called
entanglement concentration\cite{Concentration}
and the reverse operation is called entanglement dilution.

Since entanglement cannot increase under LOCC, the above
procedures are optimal, in the sense that concentration and
dilution cannot produce more copies: if they could, we would be
able to produce entangled states from nothing\footnote{It would be
like the Carnot cycle for a perpetual motion machine.}.
We can thus quantify the amount of entanglement in a state by its
entropy of entanglement.  Any state is worth that many maximally
entangled states, since it can be reversibly converted into that
many states.  We call one of these maximally entangled states an
e-bit, and shall say that other states have an entanglement of E
e-bits.  Note that this quantity is additive.  That is, if we have
two states which individually have entanglement $E_1$ and $E_2$,
together they have entanglement $E_1 + E_2$.

The quantum procedure of entanglement concentration can directly
be mapped into an equivalent classical analog. The reason for this
is that all the quantum actions used for entanglement
concentration take place in the Schmidt decomposition bases, i.e.
the unitary actions are all permutations in the Schmidt basis
while the measurements are of operators whose eigenstates are
direct products in the Schmidt basis. Hence all these actions are
essentially classical. Furthermore the quantum procedure does
not require communication, so is completely secure.

The quantum dilution protocol also has a classical analog. Indeed,
the quantum dilution \cite{Concentration} involves only Schumacher
compression
of quantum information and teleportation. Both these protocols
have classical analogs: Schumacher compression maps into Shannon
data compression and teleportation is replaced by the one-time pad
secret communication.

Since secret correlations cannot increase under LOPC, these procedures
are optimal.  They allow us to reversibly convert $N$ copies of
the classical pure state $X \sim {p_i}$ into $K$ copies of
the shared secret bit $Y \sim {q_j}$,
\begin{equation}
P( Y_A=1, Y_B = 1) = P(Y_A=2,Y_B=2) = \frac{1}{2},
\end{equation}
where
\begin{equation}
\frac{K}{N} = - \sum_i p_i log p_i.
\end{equation}
We can thus quantify the amount of secret correlations by the
entropy of secrecy, which is defined as the number of shared
secret bits which can be produced per copy of the original
state $X$.  We note that this amount is equal to the mutual
entropy between $X_A$ and $X_B$, and is also equal to the
local entropy of $X_A$, and to the local entropy of $X_B$.

\section{Entanglement Purification and Privacy Amplification}

An important procedure in quantum information is Entanglement
Purification \cite{Purification}, which turns mixed states into
pure states, at the many copy level.  The number of pure states
produced per input mixed state is the yield.

Analogous procedures for turning classical mixed states into classical
pure states exist, though are usually subdivided into two stages.
The first stage takes the mixed state $P(X_A, X_B, X_E)$ and turns
it into a mixed state where Alice and Bob hold the same value, ie.
of the form $P(i, j, k) = \delta_{ij} P(i, i, k)$. This stage is
known as Information Reconciliation \cite{Amplification}, because
Alice and Bob are agreeing on a common value.  The second stage
takes the output of the first stage, and factors out Eve, to give
a state of the form $\delta_{ij} p_i \tilde{P}(k)$. In other words
it produces a pure state.  This stage is known as Privacy
Amplification \cite{Amplification}, because Alice and Bob are
increasing the secrecy of their key by reducing (to 0) Eve's
knowledge of it.

In general it is not known what the optimal
protocol is, and there may be different optimal
protocols for different states.  There are a few different schemes
for the quantum and classical cases, but we do not wish to
discuss the details
here, just to draw the analogy.  Firstly, any
information reconciliation/privacy amplification
protocol may be used as a entanglement purification protocol.
Secondly any entanglement purification protocol may be used
as a information reconciliation/privacy amplification protocol.
We hope that a detailed study of the two problems together
will yield better understanding and new protocols in both
the classical and the quantum case.

\section{Bound Entanglement}

One of our motivations for this work was a paper\cite{Wolf} by N.
Gisin and S. Wolf suggesting a classical analog of bound
entanglement.  A bound entangled state is a bi-partite mixed
quantum state which cannot be created locally (without any prior
entanglement), but from which no maximally entangled states can be
distilled, even if there are many copies of the bound entangled
state.  It is as if the entanglement is ``bound'' inside the
state, and cannot be released.  They proposed the classical analog
to be a sample from a probability distribution on Alice, Bob and
Eve, $P(X_A, X_B, X_C)$, in which Alice and Bob have strictly
positive intrinsic information\footnote{a classical measure which,
loosely speaking, is designed to test whether or not Alice and Bob
share some information which Eve does not have and which they can
use.  The hope was that if positive, then they would have something
useful, and if zero, then they would have nothing.},
but from which they cannot
distill shared secret bits under LOPC, even if they have many
samples from the distribution. Though it is not yet known if such
a classical state exists, there is strong evidence that, by
starting with a bound entangled state $\rho_{AB}$, taking a
natural purification, $\ket{\psi_{ABE}}$, and measuring it in
natural bases, we may produce a classical bound state.  Here we
simply note that bound information fits into our framework as a
derived analogy, and is another consequence of the deeper analogy
between entanglement and secret classical correlations.

\section{Pure or Mixed?}

\label{sec-pure}

We have mentioned in section \ref{sec-states} that it is not clear
whether to classify classical states of the form $P(X_A, X_B)
P(X_E)$ where $X_A$ is {\it not} completely correlated with $X_B$
as pure or as mixed. Such a distribution resembles a pure state
because it is not correlated with Eve: this is like a pure state
not being entangled with the environment.  It also resembles a
pure state because we can optimally distill shared secret bits
from many copies of such a state at a rate equal to the natural
measure of shared correlations, the mutual information
\cite{Mutual}; this is the analog of pure state entanglement
concentration. However, it is not known whether such a distillation is
reversible. That is, given the shared secret bits, can we produce
the original states? If the answer is no, this would be typical
behavior of a mixed state. Furthermore, a definite similarity to
mixed states is that there is no Schmidt decomposition for such
states: in other words there is no way, using local reversible
transformations, to make Alice and Bob have the same values for
their samples.

Another similarity to mixed states is that it is not possible,
even probabilistically, to use LOPC to produce a pure state
 from one copy of such a distribution.
For consider the bi-partite, 2-d case, where
Alice and Bob both receive either a 0 or a 1, with probabilities
$p_{00},p_{10},p_{01},p_{11}$.  We can assume that at least the
first three probabilities are non-zero (otherwise they have a
 pure state).
They wish to use LOPC to make a classical ``entangled'' pure state,
ie. where $P(00)>0$, $P(11)>0$, $P(01)=P(10)=0$.
As discussed in section\ref{purestatesection},
the most general thing they can do is to first
communicate publicly, resulting in some total public message,
$m_i$, where i may depend upon their local dice and upon their
samples. They may then change their samples according to some map
which is specified by the message.  For example, the message could
tell Alice to flip her bit, and Bob to leave his alone.  Note that
the message has to tell them what to do locally: it cannot tell
them to look at the other person's bit to decide what they will
do.  Now, to make a pure state with any probability they need at
least one map which is local in the sense described above and
which produces both 00 and 11, and nothing else.  We shall show
that no such map exists.

Assume that such a map exists.  Without loss of generality,
we may assume the map does
\begin{equation}
00 \mapsto 00.
\end{equation}
Since Bob has to act locally, this means that if he starts
with a 0, he has to finish with a 0.  Since they must finish
with the same thing, this implies
\begin{equation}
10 \mapsto 00.
\end{equation}
Since they are symmetric, similar reasoning gives
\begin{equation}
01 \mapsto 00.
\end{equation}
Because they have to act locally, we now know that if Alice
or Bob sees a 1, they have to finish with a 0.  Thus
\begin{equation}
11 \mapsto 00.
\end{equation}
And so the map takes everything to 00, which is no good.
For classical states in higher dimensions, the same type
of reasoning shows that we cannot produce a classical
pure state from a single copy of such a state.

So, as we have shown, classical states of the form
$P(X_A,X_B)P(X_E)$ have some characteristics in common
with pure quantum states, and some in common with mixed
quantum states.

\section{Multi-Partite Results}

\label{sec-Multi}

It is well known that entanglement is much more complicated for
multi-partite systems than for bi-partite
systems\cite{multi1},\cite{multi2}, \cite{plenio}. In particular,
already in the case of three parties, it is known that tri-partite
entanglement is fundamentally different to bi-partite
entanglement, even in the many copy scenario. Furthermore, there
might even exist many different inequivalent forms of tri-partite
entanglement. As more systems are added the problem becomes vastly
more complicated, but we have a few results to guide us, such as
the fact that there is genuine entanglement at every level (again,
even in the many copy scenario). Here we show that many of these
features have classical analogs.

First, we shall look at the tripartite case.  We propose that the
classical equivalent of the GHZ state,
\begin{equation}
 \ket{GHZ}_{ABC} = \frac{1}{\sqrt{3}}(\ket{000} + \ket{111}),
\end{equation}
is a probability distribution of the form
\begin{equation}
P(X_A,X_B,X_C,X_E) = P(X_A,X_B,X_C)\tilde{P}(X_E),
\end{equation}
 where $P(X_A,X_B,X_C)$ is given by
\begin{equation}
P(0,0,0) = P(1,1,1) = \frac{1}{2}.
\end{equation}
We shall call this the C-GHZ (classical GHZ), and the classical
singlet (ie. the bipartite shared secret bit) we shall call the C-EPR.
Is is easy to see that out of 1 GHZ copy we may generate one
C-EPR, ie.
\begin{equation}
C-GHZ \mapsto C-EPR.
\end{equation}
Clare simply forgets her bit. This may sound unsatisfactory since
in the quantum case Alice and Bob end with an EPR which Clare has
no control over, whereas here Clare could always later remember
her bit, and so one may argue that we have not really performed
the classical transformation.  However, since Alice, Bob and Clare
all begin with the same information and communicate only publicly,
it is impossible for Alice and Bob to agree upon anything without
Clare knowing it.  Thus the ``stronger'' form of the
transformation is impossible, and the best we can do is this weak
form, with Clare forgetting her bit.

The above transformation is irreversible: ie. given one C-EPR, it
is impossible to make a C-GHZ\cite{multi1}.
This is because the bi-partite
entropy of secrecy can only decrease under LOPC, and viewing the
system as (AB) vs. C a $C-EPR_{AB}$ will have 0 entropy, whereas
the $C-GHZ_{ABC}$ has entropy of 1 (and is symmetric with respect
to all the parties).  It is possible, however, to do
\begin{equation}
C-EPR_{AB} + C-EPR_{BC} \mapsto C-GHZ.
\end{equation}
This is done as it would be in the quantum case: Bob makes a joint
measurement on his bits (addition modulo 2), and publicly
announces the result.  Bob now forgets his second bit, and if the
public message was 1, Clare flips her bit.  They are then done.
This procedure can be viewed as Bob using the $C-EPR_{BC}$ as a
one-time pad to send Clare the value of the $C-EPR_{AB}$.  It is
again clear that we cannot do the reverse transformation: viewing
the system as (AC) vs. B, the C-GHZ has an entropy of secrecy of
1, whereas the two C-EPR's together have an entropy of 2.

The entropy of secrecy can be used to show that there exists more
than just bi-partite secrecy, even in the many-copy case.
Specifically, the 4-party Cat state, which has distribution
$P(X_A,X_B,X_C,X_D)$ given by
\begin{equation}
P(0,0,0,0) = P(1,1,1,1) = \frac{1}{2}
\end{equation}
(where Eve factors out) cannot be converted reversibly into C-EPR
pairs.  The proof of this is exactly the proof used for the
analogous quantum problem \cite{multi1}, and is done by partitioning the
4 parties into pairs in various ways, and looking at the entropy of entanglement,
which must be asymptotically conserved under reversible transformations.

Suppose that we could reversibly convert asymptotically a single
4-party Cat state into C-EPR pairs:
$n_{AB}$ between A and B, $n_{AC}$ between A and C, etc.  Partitioning the
system into (A) vs. (BCD) we get the equation
\begin{equation}
n_{AB} + n_{AC} + n_{AD} = 1.
\end{equation}
Partitioning the system as (B) vs. (ACD), (C) vs. (ABD)
and (D) vs. (ABC) gives
\begin{equation}
n_{AB} + n_{BC} + n_{BD} = 1,
\end{equation}
\begin{equation}
n_{AC} + n_{BC} + n_{CD} = 1,
\end{equation}
\begin{equation}
n_{AD} + n_{BD} + n_{CD} = 1.
\end{equation}
On the other hand, partitioning the system as
(AB) vs. (CD), (AC) vs (BD) and (AD) vs. (BC) gives
\begin{equation}
n_{AC} + n_{AD} + n_{BC} + n_{BD} = 1,
\end{equation}
\begin{equation}
n_{AB} + n_{AD} + n_{BC} + n_{CD} = 1,
\end{equation}
\begin{equation}
n_{AB} + n_{AC} + n_{BD} + n_{CD} = 1.
\end{equation}
Summing the first 4 equations together gives
\begin{equation}
2 \sum_{all pairs} n_{ij} = 4,
\end{equation}
whilst summing together the next 3 gives
\begin{equation}
2 \sum_{all pairs} n_{ij} = 3.
\end{equation}
Thus the transformation is impossible, and the 4 party classical Cat state
really is more than just bi-partite shared secret correlations.

We thus conclude that there are different types of multi-partite
secret correlations.

\section{Conclusion}

We have described a fundamental analogy between entanglement and
secret classical correlations.  The analogy is quite simple to
state.  Both are resources, and the main objects involved in the
study of such resources have a one-to-one correspondence, as given
in the table on the first page.  Due to this basic analogy, many
derived analogies follow. In particular, we have shown that
teleportation and the one-time-pad are deeply connected, that the
concept of ``pure state'' exists in the classical domain, that
entanglement concentration and dilution are essentially classical
secrecy manipulations, and that the single copy entanglement
manipulations have such a close classical analog that the
majorization results are reproduced in the classical setting. We
have pointed out that entanglement purification is analogous to
classical privacy amplification, and hope that the search for
better protocols in the two areas can go hand in hand.  We finally
showed that, as with entanglement, one can look at multipartite
shared secret correlations, and gave a flavor of how results in
the quantum setting easily transfer into the classical world.
Despite all these useful derived analogies, our main point is the
fundamental one: entanglement and shared secret correlations are
deeply related, and one should never be viewed without the other.

We want to emphasize that by no means do we claim that quantum
entanglement is a fundamentally classical effect or that there
exists a classical explanation of entanglement. The classical
analog of entanglement is nothing more nor less than a simple
analog, and has a value of its own. On the other hand, all the
aspects of quantum entanglement which are common with the
classical analog cannot be considered to be quantum. Thus many
aspects which were hitherto considered to be genuinely quantum
lose their status.

The main thrust of this paper was to identify the common aspects
of quantum entanglement and classical secret correlations. An even
more interesting question to find those aspects which are {\it
not} common. For example, we have not found any (and believe there
is no) analog of super-dense coding. It is not the case that by
having 1 secret correlation bit and by sending 1 secret bit Alice
can transmit to Bob 2 public bits. The lack of super-dense coding
manifests itself, implicitly, also by a difference in the
quantitative descriptions of teleportation and one-time pad secret
communication: in the case of teleportation Alice has to send Bob
2 classical bits while in the one-time pad Alice has to send only
1 public bit. It is only such aspects which are not common to the
two settings which are genuinely quantum. We hope that getting rid
of those aspects which were believed to be quantum but are not,
and identifying the genuine quantum ones will lead to a better
understanding of quantum entanglement. And of secret
communication.

\section{Acknowledgements}
We thank C. H. Bennett, N. Gisin, N. Linden and S. Massar  for
very helpful discussions.

 \end{multicols}

\begin{thebibliography}{}


\bibitem{Bell} J. S. Bell, Physics 1, 195 (1964).
\bibitem{teleport} C. H. Bennett, G. Brassard, C. Crepeau, R. Jozsa,
A. Peres and W. Wootters, Phys. Rev. Lett. 70 (1993) 1895.
\bibitem{superdense} C.H.Bennett and S. Wiesner, Phys. Rev. Lett.
 69 (1992) 2881.
\bibitem{complexity} For a survey, see G. Brassard,
quant-ph/0101005.
\bibitem{qcomp} There are many reviews of this topic, eg. ``Introduction
to Quantum Computation and Information'', edited by H.-K. Lo, T.
Spiller and S.Popescu.
\bibitem{Wolf} N. Gisin, S. Wolf, quant-ph/0005042.
\bibitem{Lo2} H.-K. Lo, S. Popescu, Phys. Rev. Lett 83, 1459,
(1999).
\bibitem{Concentration} C. H. Bennett, H. J. Bernstein, S. Popescu,
B. Schumacher, Phys. Rev. A 53 (1996) 2046.
\bibitem{Unitary} A. Chefles, C. R. Gilson,
S. M. Barnett, quant-ph/0003062, quant-ph/0006106; D. Collins, N.
Linden, S. Popescu, quant-ph/0005102, to appear in Phys. Rev. A;
J. Eisert, K. Jacobs, P. Papadopolous, M. B. Plenio, Phys. Rev. A
62 (2000) 052317.
\bibitem{onetimepad} G. S. Vernam, Journal of the American Institute
for Electrical Engineers, vol. 55, pp. 109-115, 1926.
\bibitem{multi1} C. H. Bennett, S. Popescu, D. Rohrlich, J. A. Smolin,
A. V. Thapliyal, Phys. Rev. A, 63 (2001) 012307.
\bibitem{multi2} N. Linden, S. Popescu, B. Schumacher, M. Westmoreland,
quant-ph/9912039.
\bibitem{Schmidt} E. Schmidt, Math. Ann. 63, 433 (1907).
\bibitem{teleportationanalog} This analogy has also been noted by
C.H.Bennett, private communication; M. Koniorczyk, T. Kiss, J.
Janszky quant-ph/0011083.
\bibitem{Lo} H.-K. Lo, S. Popescu, quant-ph/9707038; Phys. Rev. A 63, 022301 (2001).
\bibitem{Nielsen} M. A. Nielsen, Phys. Rev. Lett. 83 (1999)
436-439.

\bibitem{Wehrl} General Properties of entropy, A. Wehrl, Rev. Mod. Phys.
Vol 50, No. 2 (1978) 221-260.
\bibitem{Vidal} G. Vidal, Phys. Rev. Lett. 83 (1999) 1046-1049.
\bibitem{Jonathan} D. Jonathan, M. B. Plenio, Phys. Rev. Lett. 83 (1999)
3566.
\bibitem{Purification} C. H. Bennett, G. Brassard, S. Popescu,
B. Schumacher, J. A. Smolin, W. K. Wootters, Phys. Rev. Lett. 76
(1996) 722-725.
\bibitem{Amplification} C.H. Bennett, C. Crepeau, U. M. Maurer, IEEE Trans.
Inform. Theory, Vol 41, pp 1915-1923, Nov 1995 and references
therein.
\bibitem{Mutual} R. Ahlswede and I. Csiszar, IEEE Trans. Inform. Theory,
vol 39, pp 1121-1132, July 1993; U. Maurer, IEEE Trans. Inform.
Theory, vol39, pp 733-742, May 1993.
\bibitem{plenio} E. F. Galvao, M. B. Plenio and S. Virmani, J. Phys. A 33, 8809
(2000).

\end{thebibliography}
\end{document}